\documentclass[a4paper]{revtex4}
\usepackage{epsfig,amsmath,amssymb}
\def\q{q \bar q}
\def\gs{\gamma_s}

\def\be{\begin{equation}}
\def\ee{\end{equation}}

\def\NP{{ Nucl.\ Phys.\ }}
\def\PL{{ Phys.\ Lett.\ }}
\def\PR{{ Phys.\ Rev.\ }}

\def\ZP{{ Z.\ Phys.\ }}
\def\EP{{ Eur.\ Phys.\ J.\ C}}

\begin{document}

~~\vskip0.7cm

\centerline{\Large \bf Hawking-Unruh Hadronization}
\bigskip
\centerline{\Large \bf and}
\bigskip

\centerline{\Large \bf Strangeness Production In High Energy Collisions}

\vskip0.5cm 

\centerline{\bf Paolo Castorina$^{\rm a,b,c}$ and Helmut Satz$^{\rm d}$} 

\bigskip
\centerline{a: Dipartimento di Fisica ed Astronomia, Universita' di Catania, Italy}
\centerline{b: INFN sezione di Catania, Catania, Italy}
\centerline{c: PH Department, TH Unit, CERN, CH-1211 Geneva 23, Switzerland}
\centerline{d: Fakult\"at f\"ur Physik, Universit\"at Bielefeld, Germany}



\vskip1cm

\centerline{\large \bf Abstract}

\bigskip

The thermal multihadron production observed in different high energy 
collisions poses many basic problems: why do even elementary,
$e^+e^-$ and hadron-hadron, collisions show thermal behaviour? Why is there in such interactions a  suppression of strange particle production? Why does the strangeness suppression almost disappear in relativistic heavy ion collisions? Why in these collisions is the thermalization time less than $\simeq 0.5 $ fm/c?

We show that the recently proposed mechanism of thermal hadron production through Hawking-Unruh radiation can naturally answer the previous questions.
Indeed,  the interpretation of quark ($q$)- antiquark ($\bar q$) pairs production, by the sequential string breaking,  as tunneling through  the event horizon of colour confinement leads to thermal behavior with a universal temperature, $T \simeq 170$ Mev,related to the quark acceleration, $a$, by $T=a/2\pi$. The resulting temperature depends on the quark mass and then on the  content of the produced hadrons,  causing a deviation from full equilibrium and hence a suppression of strange particle production in elementary collisions. In nucleus-nucleus collisions, where the quark density is much bigger, one has to introduce an average temperature (acceleration) which dilutes the quark mass effect and the strangeness suppression almost disappears.

\vskip1cm

\section{Introduction}

Hadron production in high energy collisions shows remarkably universal thermal 
features. In $e^+e^-$ annihilation \cite{Beca-e,erice,Beca-h}, in $pp$, 
$p\bar p$ \cite{Beca-p} and more general $hh$ interactions \cite{Beca-h}, 
as well as in the collisions of heavy nuclei \cite{Beca-hi}, over an energy range from
around 10 GeV up to the TeV range, the relative abundances of the produced 
hadrons appear to be those of an ideal hadronic resonance gas at a quite
universal temperature $T_H \approx 160-170$ MeV \cite{Beca-Biele}. 

\medskip

There is, however, one important non-equilibrium effect observed:
the production of strange hadrons in elementary collisions is suppressed 
relative to an overall equilibrium. This is usually taken into account 
phenomenologically by introducing an overall strangeness suppression factor 
 $\gs < 1$ \cite{Raf}, which reduces the predicted abundances by $\gs, 
\gs^2$ and $\gs^3$ for hadrons containing one, two or three strange quarks 
(or antiquarks), respectively. In high energy heavy ion collisions, 
strangeness suppression becomes less and disappears at high energies 
\cite{gamma}.

\medskip

The origin of the observed thermal behaviour has been an enigma for many years 
and there is a still ongoing debate about the interpretation of these
results \cite{diba}. Indeed, in high energy heavy ion collisions multiple parton 
scattering could lead to kinetic thermalization, but $e^+e^-$ or elementary hadron interactions do not readily allow such a description. 

The universality of the observed temperatures, on
the other hand, suggests a common origin for all high energy collisions, 
and it was recently proposed \cite{CKS} that thermal hadron production is the 
QCD counterpart of Hawking-Unruh (H-U) radiation \cite{Hawk,Un}, emitted at the
event horizon due to colour confinement. 
\medskip
In the case of approximately massless quarks, the resulting 
hadronization temperature is determined by the string tension $\sigma$, with 
$T \simeq \sqrt{\sigma/2\pi} \simeq 170$ Mev  \cite{CKS}. Moreover in ref.\cite{noi} it has been shown that  strangeness suppression in elementary collisions naturally occurs  in this framework  without requiring a specific suppression factor. The crucial role here is played  by the non-negligible strange quark mass, which modifies the emission temperature for such quarks. 

\medskip

In this work  we briefly review previous results and show that when the quark density is much bigger than in elementary scattering, as in relativistic heavy ion collisions,
the effect of strange quark mass is washed out by the average acceleration due to the large number of light quarks and the strangeness suppression disappears.

In the second section, after recalling the statistical hadronization model, one discusses  the subsequent description in terms of H-U radiation in QCD, including the  dependence of the radiation temperature on the mass of the produced quark and the strangeness suppression in elementary collisions. In Sec.3 the dynamical mechanism for strangeness enhancement (i.e. no suppression) in relativistic heavy ion scatterings is introduced giving a coarse grain evaluation of the Wroblewski factor \cite{wrob}, a parameter which counts the relative number of strange and non strange quarks and antiquarks. Sec.4 is devoted to comments and conclusions.

\section{Hawking-Unruh Hadronization}

In this section we will recall the essentials of the statistical 
hadronization model and of the Hawking-Unruh analysis of the string breaking mechanism . For a detailed descriptions see ref.~\cite{Beca-h,CKS,noi}.

\medskip

\subsection{ Statistical hadronization model}

The statistical hadronization model assumes that hadronization in high 
energy collisions is a universal process proceeding through the formation of 
multiple colourless massive clusters (or fireballs) of finite spacial 
extension. These clusters are taken to decay into hadrons according 
to a purely statistical law: every multi-hadron state of the cluster phase space 
defined by its mass, volume and charges is equally probable.
The mass distribution and the distribution of charges (electric, baryonic 
and strange) among the clusters and their (fluctuating) number are 
determined in the prior dynamical stage of the process. 
Once these distributions are known, each cluster can be hadronized on
the basis of statistical equilibrium, leading to the calculation of 
averages in the {\em microcanonical ensemble}, enforcing the exact 
conservation of energy and charges of each cluster. 

\medskip

Hence, in principle, one would need the mentioned dynamical information 
in order to make definite quantitative predictions to be compared with 
data. Nevertheless, for Lorentz-invariant quantities such as multiplicities, 
one can introduce a simplifying assumption and thereby obtain a simple 
analytical expression in terms of a temperature. The key point is to assume
that the distribution of masses and charges among clusters is again
purely statistical \cite{Beca-h}, so that, as far as the calculation 
of multiplicities is concerned, the set of clusters becomes equivalent, 
on average, to a large cluster ({\em equivalent global cluster}) whose 
volume is the sum of proper cluster volumes and whose charge is the sum of 
cluster charges (and thus the conserved charge of the initial colliding 
system). In such a global averaging process, the equivalent cluster 
generally turns out to be large enough in mass and volume so that the 
canonical ensemble becomes a good approximation. In other words, a temperature 
can be introduced which replaces the {\sl a priori} more fundamental 
description in terms of an energy density. 

To obtain a simple expression for our further discussion, we neglect
for the moment an aspect which is important in any actual analysis. 
Although in elementary collisions the conservation of the various discrete 
Abelian charges (electric charge, baryon number, strangeness, heavy flavour) 
has to be taken into account {\sl exactly} \cite{HR},  we here consider for the moment 
a grand-canonical picture. We also assume Boltzmann distributions for all 
hadrons. The multiplicity of a given scalar hadronic species $j$ then becomes

\be
\langle n_j \rangle^{\rm primary} = \frac{V T m_j^2}{2\pi^2} 
\gamma_s^{n_j} {\rm K}_2\left(\frac{m_j}{T}\right)\, 
\ee
with $m_j$ denoting its mass and $n_j$ the number of strange quarks/antiquarks
it contains. Here primary indicates that it gives the number at the
hadronisation point, prior to all subsequent resonance decay.
The Hankel function $K_2(x)$, with $K(x) \sim 
\exp\{-x\}$ for large $x$, gives the Boltzmann 
factor, while $V$ denotes the overall equivalent cluster volume. In other 
words, in an analysis of $4 \pi$ data of elementary collisions, $V$ is the 
sum of the all cluster volumes at all different rapidities. It thus scales
with the overall multiplicity and hence increases with collision energy.  
A fit of production data based on the statistical hadronisation model thus
involves three parameters: the hadronisation temperature $T$, the 
strangeness suppression factor $\gamma_s$, and the equivalent global cluster
volume $V$. 

As previously discussed, at high energy the temperature turns out to be independent  on the initial configuration and this result calls for a universal mechanism underlying the hadronization. In the next paragraph we recall the interpretation of the string breaking as  QCD H-U radiation.

\subsection{String breaking and event horizon}

Let us outline the thermal hadron production process through
H-U radiation for the specific case of $e^+e^-$ annihilation
(see Fig.\ \ref{breaking}). The separating primary $\q$ pair excites a 
further pair $q_1\bar q_1$ from the vacuum, and this pair is in turn pulled 
apart by the primary constituents. In the process, the $\bar q_1$ shields 
the $q$ from its original partner $\bar q$, with a new $q\bar q_1$ string 
formed. When it is stretched to reach the pair production threshold, a 
further pair is formed, and so on \cite{bj,nus}. Such a pair production mechanism is a special case of H-U radiation \cite{KT},  emitted as hadron $\bar q_1 q_2$  when the quark $q_1$ tunnels through its 
event horizon to become $\bar q_2$.  
\begin{figure}[h]
\centerline{\psfig{file=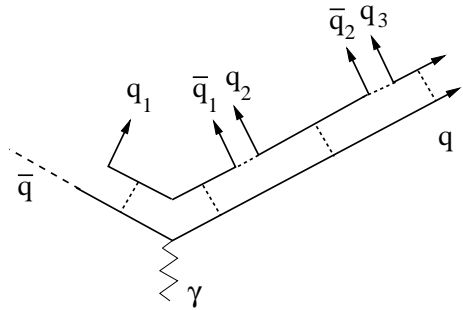,width=6cm}} 
\caption{String breaking through $\q$ pair production}
\label{breaking}
\end{figure}
The corresponding hadron radiation temperature is given by the Unruh form 
$T_H = {a/2 \pi}$, where $a$ is the acceleration suffered by the quark 
$\bar q_1$ due to the force of the string attaching it to the primary quark 
$Q$. This is equivalent to that suffered by quark $q_2$ due to the 
effective force of the primary antiquark $\bar Q$. Hence we have
\be
a_q = {\sigma \over w_q} =
{\sigma \over \sqrt{m_q^2 + k_q^2}}, 
\ee
where $w_q =\sqrt{m_q^2 + k_q^2}$ is the effective mass of the produced 
quark, with 
$m_q$ for the bare quark mass and $k_q$ the quark momentum inside the 
hadronic system 
$q_1\bar q_1$ or $q_2\bar q_2$. Since the string breaks \cite{CKS} when 
it reaches 
a separation distance 
\be
x_q \simeq {2\over \sigma} \sqrt{m^2_q + (\pi \sigma /2)},
\ee
the uncertainty relation gives us with $k_q \simeq 1/x_q$
\be
w_q  = \sqrt{m_q^2 + [\sigma^2/(4m_q^2 + 2\pi \sigma)]} 
\ee
for the effective mass of the quark. The resulting quark-mass dependent 
Unruh temperature is thus given by
\be
T(qq) \simeq {\sigma \over 2\pi w_q}.
\label{Tq}
\ee
Note that here it is assumed that the quark masses for $q_1$ and $q_2$
are equal. For $m_q \simeq 0$, eq.\ (\ref{Tq}) reduces to
\be\label{T0}
T(00) \simeq \sqrt{\sigma \over 2\pi},
\ee
as obtained in \cite{CKS}.

\medskip

If the produced hadron ${\bar q}_1 q_2$ consists of quarks of different 
masses, the resulting temperature has to be calculated as an average
of the different accelerations involved. For one massless quark 
($m_q \simeq 0$) and one of strange quark mass $m_s$, the average
acceleration becomes
\be
\bar a_{0 s} = {w_0 a_0 + w_s a_s \over w_0 + w_s} = 
{2\sigma \over w_0 + w_s}.
\ee
From this the Unruh temperature of a strange meson is given by
\be\label{T0s}
T(0s) \simeq {\sigma \over \pi (w_0 + w_s)},
\ee
with $w_0 \simeq \sqrt{1/2\pi\sigma}$ and $w_s$ given by eq.(4) with $m_q=m_s$. Similarly, we obtain
\be\label{Tss}
T(ss) \simeq {\sigma \over 2 \pi w_s},
\ee
for the temperature of a meson consisting of a strange quark-antiquark pair 
($\phi$).  
With $\sigma\simeq 0.2$ GeV$^2$, eq.\ (\ref{T0}) gives $T_0 \simeq 0.178$ 
GeV. A strange quark mass of 0.1 GeV reduces this to $T(0s) \simeq 0.167$ 
GeV and $T(ss)\simeq 157$ MeV, i.e., by about 6 \% and 12 \%, respectively.

\medskip

The scheme is readily generalized to baryons. The production
pattern is illustrated in Fig.\ \ref{baryon} and leads to an
average of the accelerations of the quarks involved. We thus have
\be
T(000) = T(0) \simeq {\sigma \over 2\pi w_0}
\ee
for nucleons, 
\be
T(00s) \simeq {3 \sigma \over 2\pi(2w_0 + w_s)}
\ee
for $\Lambda$ and $\Sigma$ production, 
\be
T(0ss) \simeq {3 \sigma \over 2\pi(w_0 + 2w_s)}
\ee
for $\Xi$ production, and 
\be
T(sss) = T(ss) \simeq {\sigma \over 2\pi w_s}
\ee
for that of $\Omega$'s. 
We thus obtain a resonance gas picture with five different hadronization 
temperatures, as specified by the strangeness content of the hadron in
question: $T(00)=T(000),~T(0s),~T(ss)=T(sss),~T(00s)$ and $T(0ss)$.

\begin{figure}[h]
\centerline{\psfig{file=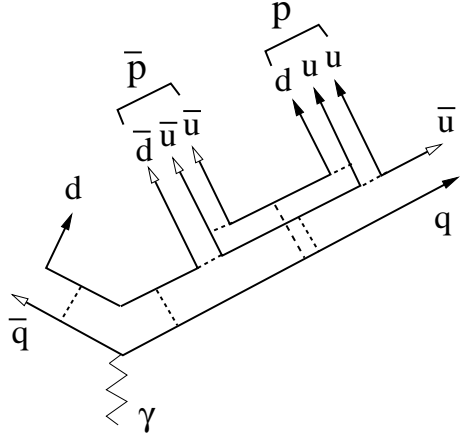,width=6cm}} 
\caption{nucleon formation by string breaking through $\q$ pair production}
\label{breaking}
\end{figure}

\subsection{Strangeness suppression in elementary collisions}

In order to evaluate the primary hadron multiplicities, the previous different species-dependent temperatures, fully determined by the string tension $\sigma$ and the strange quark mass $m_s$,  have been inserted into a complete statistical model calculation \cite{noi}. 

Apart from possible variations of the quantities of $\sigma$ and $m_s$, the description is thus parameter-free. As illustration, we show in table \ref{tab:1} the 
temperatures obtained for $\sigma = 0.2$ GeV$^2$ and three different 
strange quark masses. It is seen that in all cases, the temperature for 
a hadron carrying non-zero strangeness is lower than that of non-strange 
hadrons and this leads to an overall strangeness suppression in elementary collisions, in good agreement with data \cite{noi}, without the introduction of the ad-hoc parameter $\gamma_s$. 

\begin{table}[h]
\begin{center} 
\begin{tabular}{|c|c|c|c|}  \hline
\hline
$ T  $ & $m_s =0.075$ & $ m_s=0.100 $ & $m_s= 0.125 $ \\
\hline
$T(00)$ & 0.178  & 0.178  &  0.178 \\
$T(0s)$ & 0.172  & 0.167  &  0.162 \\
$T(ss)$ & 0.166  & 0.157  &  0.148 \\
$T(000)$& 0.178  & 0.178  &  0.178 \\
$T(00s)$& 0.174  & 0.171  &  0.167 \\
$T(0ss)$& 0.170  & 0.164  &  0.157 \\
$T(sss)$& 0.166  & 0.157  &  0.148 \\
\hline
\end{tabular}
\caption{Hadronization temperatures for hadrons of different strangeness 
content, for $m_s= 0.075 , 0.100,0.125$ Gev and $\sigma=0.2$ GeV$^2$.}
\label{tab:1}
\end{center}
\end{table}

\medskip

Although a complete statistical hadronization model calculation has been necessary for a detailed comparison with experimental results \cite{noi}, a rather simple, coarse grain, argument can illustrate why a reduction of the H-U temperature for strange particle production reproduces the $\gamma_s$ effect.

To simplify matters, let us assume that there are only two species: scalar and electrically neutral mesons, "pions" with mass $m_\pi$, and "kaons" with mass $m_k$ and strangeness $s=1$.

According to the  statistical model with the $\gamma_s$ suppression factor, the ratio $N_k/N_\pi$ is obtained by eq.(1) and is given by
\be
\frac{N_k}{N_\pi}|^{stat}_{\gamma_s} = \frac{m_k^2}{m_\pi^2} \gamma_s \frac{K_2(m_k/T)}{K_2(m_\pi/T)}
\ee
because there is thermal equilibrium at temperature $T$.

On the other hand, in the H-U based statistical model there is no $\gamma_s$, $T_k=T(0s) \ne T_\pi=T(00)=T$  and therefore
\be
\frac{N_k}{N_\pi}|^{stat}_{H-U} = \frac{m_k^2}{m_\pi^2} \frac{T_k}{T_\pi} \frac{K_2(m_k/T_k)}{K_2(m_\pi/T_\pi)}.
\ee
 which corresponds to a $\gamma_s$ parameter given by
\be
\gamma_s=  \frac{T_k}{T_\pi} \frac{K_2(m_k/T_k)}{K_2(m_k/T_\pi)}.
\ee
For $\sigma=0.2$ Gev$^2$, $m_s=0.1$ Gev, $T_\pi = 178$ Mev and $T_k=167$ Mev ( see table I), the crude evaluation by eq.(16) gives $\gamma_s \simeq 0.73$.

\medskip

In conclusion, our picture implies that the produced hadrons are emitted slightly ``out of equilibrium'', in the sense that the emission temperatures are not identical. 
As long as there is no final state interference between the produced quarks 
or hadrons, we expect to observe this difference and hence a modification 
of the production of strange hadrons, in comparison to the corresponding 
full equilibrium values. 

Once such interference becomes likely, due to the increases of parton density ,as in high energy heavy ion collisions, equilibrium can be essentially restored producing a small strangeness suppression as we shall discuss in the next section.

\section{Hawking-Unruh strangeness enhancement in heavy ion collisions}

As previosuly discussed, in elementary collisions the production of hadrons containing $n$ strange quarks or antiquarks is reduced in
comparison to nucleu-nucleus scattering and this reduction can be
accounted for by the introduction of the phenomenological universal strangeness suppression factor $\gamma_s^n$.

However, the hadron production in high energy collisions occurs  in a number of causally disconnected regions of finite space-time size \cite{noi2}. As a result, globally conserved quantum numbers (charge, strangeness, baryon number) must be conserved locally in spatially restricted correlation clusters. This provides a dynamical basis for understanding  the  suppression  of strangeness production in elementary interactions ($pp$, $e^+e^-$) due to a small strangeness correlation volume \cite{HR,BRSt,kraus1,kraus2}. 

The H-U counterpart of this mechanism is that in elementary collisions there is a small number of partons in a causally connected region and the hadron production comes from the sequential breaking of independent $q \bar q$ strings with the consequent species-dependent temperatures which reproduce the strangeness suppression.

In contrast, the space-time superposition of many collisions in heavy ion 
interactions largely removes these causality constraints \cite{noi2}, resulting in 
an ideal hadronic resonance gas in full equilibrium. 

In the H-U approach, the effect of a large number of causally connected quarks and antiquarks can be implemented by defining the average temperature of the system and determining the hadron multiplicities by the statistical model with this "equilibrium" temperature.

Indeed, the average temperature depends on the numbers of light quarks, $N_l$, and of strange quarks, $N_s$, which, in turn, are counted by the number of strange and non-strange hadrons in the final state at that temperature.

A detailed analysis requires again a full calculation in the statistical model, that will be done in a forthcoming paper, however the mechanism can be roughly illustrated in the world of "pions" and "kaons" introduced in sec.2.

Let us consider a high density system of quarks and antiquarks in a causally connected region. Generalizing our formulas in sec. 2, the average acceleration is given by
\be
\bar a = \frac{N_l w_0 a_0 + N_s w_s a_s}{N_l w_0 + N_s w_s}
\ee
By assuming $N_l >> N_s$, by eqs.(6-8), after a simple algebra, the average temperature, $\bar T = \bar a/2\pi$,  turns out to be
\be
\bar T = T(00)[ 1 - \frac{N_s}{N_l} \frac{w_0 + w_s}{w_0} (1 - \frac{T(0s)}{T(00)})]
 + O[(N_s/N_l)^2]
\ee
Now in our world of "pions" and "kaons" one has $N_l = 2 N_\pi + N_k$ and $N_s=N_k$ and therefore
\be
\bar T = T(00)[ 1 - \frac{N_k}{2 N_\pi} \frac{w_0 + w_s}{w_0} (1 - \frac{T(0s)}{T(00)})]
 + O[(N_k/N_\pi)^2].
\ee
On the other hand, in the H-U based statistical calculation the ratio $N_k/N_\pi$ depends on the equilibrium (average) temperature $\bar T$, that is
\be
N_k/N_\pi = \frac{m_k^2}{m_\pi^2} \frac{K_2(m_k/\bar T)}{K_2(m_\pi/\bar T)},
\ee 
and, therefore, one has to determine the temperature $\bar T$ in such a way that eq.(19) and eq.(20) are self-consistent. This condition implies the equation
\be
2 \frac{[1 - \bar T/T(00)]w_0}{[1-T(0s)/T(00)](w_s+w_0)} =  \frac{m_k^2}{m_\pi^2} \frac{K_2(m_k/\bar T)}{K_2(m_\pi/\bar T)},
\ee
that can be solved numerically.

For $\sigma=0.2$ Gev$^2$, $m_s=0.1$ and the temperatures in  table I, the average temperature turns out $ \bar T = 174$ Mev and one can evaluate the Wroblewski factor defined by
\be
\lambda = \frac{2N_s}{N_l}
\ee
where $N_s$ is the number of strange and anti-strange quarks in the hadrons in the final state and $N_l$ is the number of light quarks and antiquarks in the final state    minus their number in the initial configuration. 

The experimental value of the Wroblewski factor in high energy collisions is rather independent on the energy and is about $\lambda \simeq 0.26$ in elementary collisions and $\lambda \simeq 0.5$ for nucleus-nucleus scattering.

In our simplified world, for $e^+e^-$ annhilation, $\lambda = N_k/N_\pi$ and is given by eq.(15) with the species-dependent temperatures in table I.
One gets $\lambda \simeq 0.26$.

To evaluate the Wroblewski factor in nucleus-nucleus collisions one has to consider the average "equilibrium" temperature $\bar T$ and the number of light quarks in the initial configuration. The latter point requires a realistic calculation in the statitical model which includes all resonances and stable particles.

However to show that one is on the right track, let us neglect the problem of the initial configurationa and let us evaluate the effect of substituting in eq.(15) the species-dependent temperatures with the equilibrium temperature $\bar T$. With this simple modification one has an increasing of the Wroblewski factor, $\lambda = 0.33$. 

In other terms, the change from a non-equilibrium condition, with species-dependent temperatures, to an equilibrated system with the average temperature $\bar T$ is able to reproduce part of the observed growing of the number of strange quarks with respect to elementary interactions.

\section{Conclusion and Comments}

The Hawking-Unruh hadronization mechanism gives a dynamical basis to the universality of the temperature obtained in the statistical model, to the strangeness suppression in elementary collisions and to its enhancement in nucleus nucleus scatterings.

Moreover it is able to understand the extremely fast equilibration of hadronic matter.
Indeed, Hawking-Unruh radiation provides a stochastic rather than kinetic approach to equilibrium, with a randomization essentially due to the quantum physics of the color horizon. The barrier to information transfer due the event horizon requires that
the resulting radiation states excited from the vacuum are distributed according to maximum entropy, with a temperature determined by the strength of the “confining” field.
The ensemble of all produced hadrons, averaged over all events, then leads to the same
equilibrium distribution as obtained in hadronic matter by kinetic equilibration. In the case of a very high energy collision with a high average multiplicity already one event can provide such equilibrium; because of the interruption of information transfer at each of the successive quantum color horizons, there is no phase relation between two successive production steps in a given event. The destruction of memory, which in kinetic equilibration is achieved through sufficiently many successive collisions, is here automatically provided by the tunnelling process.
So the thermal hadronic final state in high energy collisions is obtained through a
stochastic process. 
  
Finally, high energy particle physics, and in particular hadron production, is, in our opinion, the promising sector to find the analogue of the Hawking-Unruh radiation for two main reasons: color confinement and the huge acceleration that cannot be reached in any other dynamical systems.


\end{document}